# A REVIEW OF THE RECENT DEVELOPMENTS IN THE FABRICATION PROCESSES OF CMOS IMAGE SENSORS FOR SMARTPHONES

Kirthika Nahalingam, Linda P. B. Katehi
*Department of Electrical and Computer Engineering*
*Texas A&M University, College Station, TX, USA.*
kirthika@tamu.edu, katehi@tamu.edu

*Abstract*—CMOS Image Sensors are experiencing significant growth due to their capabilities to be integrated in smartphones with refined image quality. One of the major contributions to the growth of image sensors is the innovation brought about in their fabrication processes. This paper presents a detailed review of the different fabrication processes of the CMOS Image Sensors and its impact on the image quality of smartphone pictures. Fabrication of CMOS image sensors using wafer bonding technologies such as Through-Silicon-Vias and Cu-Cu hybrid bonding along with their experimental results are discussed. A 2-layer architecture of photodiode and pixel transistors has adopted the 3D sequential integration, by which the wafers are bonded together one after the other in the fabrication process. Electrical characteristics and reliability test results are presented for the former two fabrication processes and the improvements in the pixel's performance such as conversion gain, quantum efficiency, full well capacity and dynamic range for the 2-layer architecture are discussed.

## I. Introduction

Cameras are one of the many features that has contributed to the significant growth of the smartphones industry. In the recent years, image quality of the camera phones and smartphones have improved greatly, such that the market share for digital cameras have declined [1] and this is shown in the Fig. 1. This is due to the capability to miniaturize and embed image sensors in the smartphones and being able to get the image quality on par with the digital cameras.

Cameras have a long history since the 19$^{th}$ century. The first digital camera prototype was made in the year 1973 at Bell Labs and it was based on the image sensor called Charge Coupled Device (CCD) [2]. Later, the first commercial digital camera, Cromemco Cyclops based on the MOS sensor was introduced in the year 1975. This camera did not have a permanent storage of its own and had to be connected to a computer to retrieve the image [2]. Fig. 2 shows the 32x32 pixel image photographed by the Cyclops.

Most of the smartphone cameras available today are based on CMOS image sensor (CIS) technology. The industries began to adopt the CIS technology for the smartphone cameras since their invention at the Jet Propulsion Laboratory in 1993 [4]. Since then, the image quality of the smartphone pictures have improved greatly due to the developments in CIS fabrication process. This is best understood from Fig. 3 which shows the images taken by iPhones (with CIS) more than a decade ago. The improvement in the picture quality is marked by the yellow box which both the iPhone 4 (2010) and iPhone 3GS (2009) captured with minimum brightness. The latest boost in the CIS can be seen in the picture taken by iPhone 14 (2022) which was shot in a dark sky as seen in the Fig. 4. The growth of CIS and its contribution in the smartphone industry is evident comparing the images of Fig. 3 and Fig. 4.

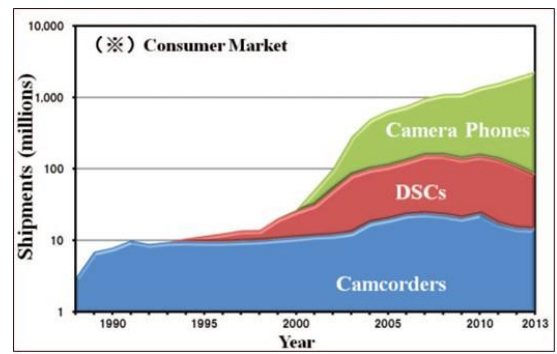

Fig. 1. Year vs Device Shipments [1]

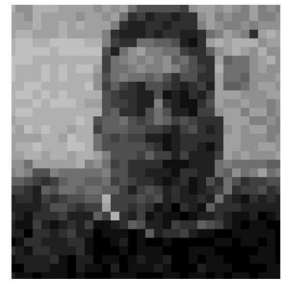

Fig. 2. Image taken by Cromemco Cyclops [3]

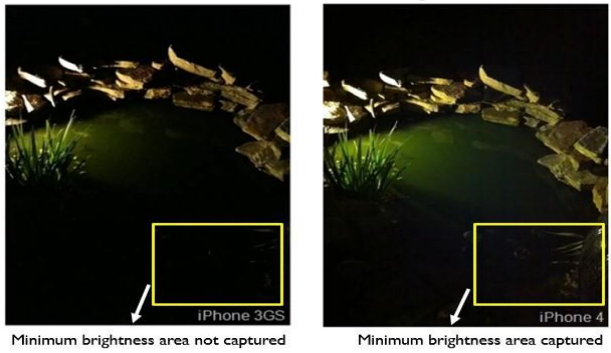

Fig. 3. Images taken by 2009 iPhone3GS (left) and 2004 iPhone4 (right) Modified [5]

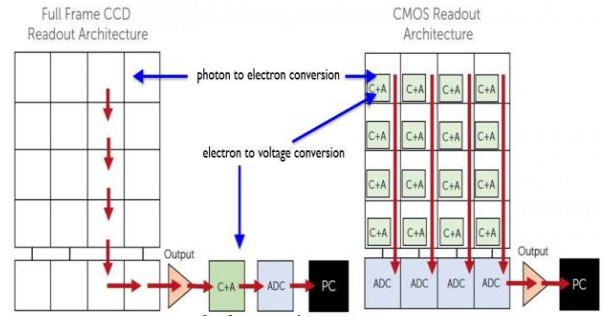

Fig. 5. CCD vs CIS Modified [8]

*B. Evolution of CIS*

The CMOS image sensor available in the market today has evolved since its invention in 1993. The conventional CIS is front illuminated, which has the microlens and color filter at the top, followed by the metal wiring for interconnects and the photodiode at the bottom of the architecture. Since the light enters the image sensor through metal layers, some of the light information is reflected and lost during its transit before even reaching the photodiode. Clearly, this arrangement affected the front-illuminated sensor's performance and the issue was overcome when Sony Corporation came up with the idea of moving the photodiode to the top of the architecture next to the color filter. This architecture is known as the back-illuminated (BI) CMOS image sensor and it improved the sensor's performance greatly, thereby marking the beginning of a new era for the CMOS image sensors [9]. A comparison of the front and back illuminated CMOS image sensor can be seen in the Fig. 6.

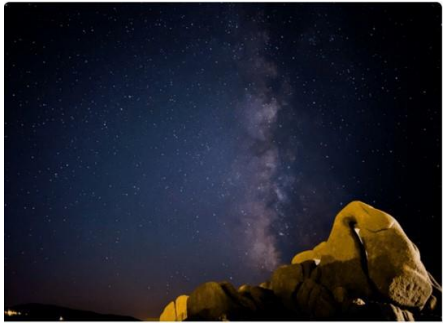

Fig. 4. Image taken by 2022 iPhone14 [6]

*A. Working Principle of CIS*

Image sensors are semiconductor devices that detect and convert photons into electrical signals which can later be converted to a meaningful image. The photons are captured and converted to electrons by the photodiode, which are then converted to voltage. This analog information is converted to digital information using an analog-to-digital converter (ADC) and further processed to get the final image [7]. Although, CIS are the extensively used image sensing technology today, their basic working principle has been adopted from its predecessor, Charge Coupled Devices.

A comparison of the CCD and CIS working operation is shown the Fig. 5. Both the CCD and CIS are comprised of pixel arrays. Each pixel is composed of a photodiode and an electrode in the CCD and a photodiode and transistor in the CIS. In CCD, when the photons hit the photodiode, they are converted to electrons and then the electrons are transferred from one pixel to another to the register at the end, where they are converted to voltage and amplified. In contrast to CCD, in a CIS pixel, the voltage information is obtained in the pixel itself and subsequently, the voltages are read at the same time using vertical or horizontal bars from all the pixels. Some CIS have ADCs on the chip itself and the output exiting the sensor is already digitized. Since the information is readout faster, the loss of information is minimal and power consumption is low in CIS, they have superseded CCD in the past decade.

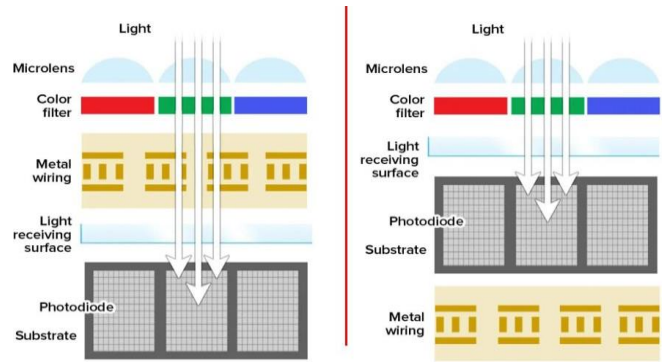

Fig. 6. Front (left) vs Back (right) illuminated CIS Modified [10]

Following the back illuminated sensor, the idea to stack the pixel and the logic circuit sections was proposed to reduce the size of the sensor in the X and Y directions. Fig. 7 shows the three dimensional view of the conventional BI-CIS (left) where the pixel and the logic circuits of the sensor share the same substrate and stacked sensor (right) where, the pixel section containing the photodiodes was placed on the top and the logic circuitry was moved to the bottom of the architecture on the supporting substrate. This is called the Stacked Back-

Illuminated CMOS Image Sensor, which was proposed by Sony Corporation in the year 2012 [10].

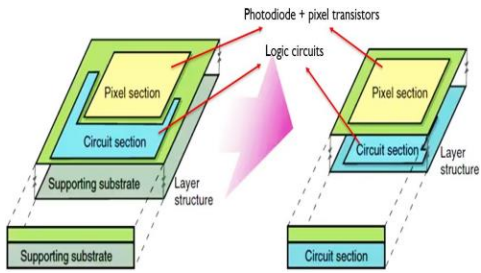

Fig. 7. Conventional BI-CIS (left) vs Stacked BI-CIS (right) illuminated CIS Modified [10]

This paper presents the various research works that was carried out by the industry and showed several improvements in the Stacked Back-Illuminated CMOS Image Sensor (BI-CIS). The research works focused on exploring various fabrication processes for the stacked sensors and on improving the different performance metrics of the photodiodes. These advancements are critical for the growth of the CMOS image sensors, as they in turn contribute to the smartphone industry.

## II. Stacked BI CIS using Through Silicon Vias, H. Tsugawa et al, 2017

Rolling shutter distortion is a common phenomenon found in both smartphones and professional digital cameras. This image distortion occurs when the object of interest to be photographed moves at a speed greater than the sensor's speed and the bottom of the frame is captured later than the top of the frame. This lag creates a distorted image. Mechanical shutters in digital cameras overcome this problem by exposing the sensor only to a specific time frame that the sensor can process. But, the mechanical shutters cannot be used in smartphones to reduce image distortion. The study probing to reduce the image distortion in smartphone CIS is the motivation behind the research work discussed in [11].

According to [11], image distortion in CMOS sensors is caused by the low reading speed at the output interface of the sensor architecture. As illustrated in Fig. 8, in the conventional stacked CIS, the image information from the pixel layer travels through the DRAM layer to the logic circuit and exits the output interface (I/F) in a serial fashion. Even though the pixel section is capable of capturing the entire image, the interface can only readout the information serially. This gives rise to the delay in the reading speed at the I/F compared to the pixel end, hence leading to the distortion. In an attempt to eliminate the distortion, adding a temporary memory to store the pixel information seemed like a viable solution. So adding a capacitor inside the pixel was implemented in [12]. But, that meant that the pixel size is increasing, thereby affecting the picture resolution.

The image distortion at the I/F was improved by making changes to the data flow in the image sensor architecture, rather than to the pixel. A 1Gbit DRAM memory layer was added in between the logic and the I/F layers and Fig. 9 shows the 3-layer pixel/DRAM/logic configuration. The output from the pixel reaches the logic layer through DRAM and then the data from the logic layer is stored in the DRAM layer momentarily until the I/F is ready to read the information. It is critical to note that the reading speed is increased to 120fps at the pixel side, while the reading speed remains 30fps at the I/F, which is the same as without the DRAM. The addition of DRAM layer and 120fps reading speed at the pixel output diminishes the rolling shutter distortion effect to a great extent in the final image.

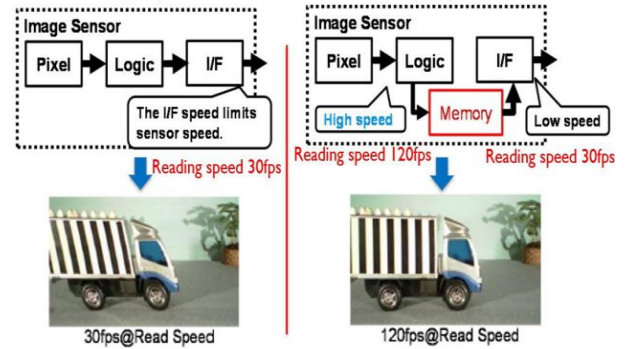

Fig. 8. Stacked BI-CIS without (left) and with (right) DRAM (Modified [11])

### A. Structure of the Stacked BI-CIS using Through Silicon Via

The structure of the 3-layer stacked CIS displayed in Fig. 9 is fabricated using the 90nm CMOS process for the pixel, 30nm for the DRAM and 40nm for the logic layer using two tiers of stacked through-silicon vias (TSV). The middle DRAM and the logic layers are connected by the lower TSV stack and the pixel substrate and lower TSV stack are connected by the upper TSV stack.

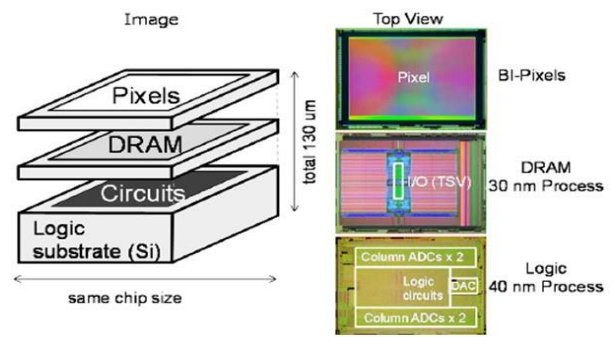

Fig. 9. Structure of 3-layer pixel/DRAM/logic CIS [11]

### B. Process Flow of Through Silicon Via

The process flow of stacking the 3-layer CIS is illustrated in the Fig. 10. The fabrication process begins with the parallel processing of wafers where each wafer is bonded to their respective substrates individually. The DRAM is flipped bonded to the logic substrate face-face. The DRAM substrate is thinned to about 3μm after the bonding. Then, the lower TSVs and the metal wiring connecting both the substrates are formed. Later, the pixel substrate is flipped and bonded to the already stacked DRAM/logic substrate and then the upper TSVs are made to connect the pixel substrate to the rest of the stack.

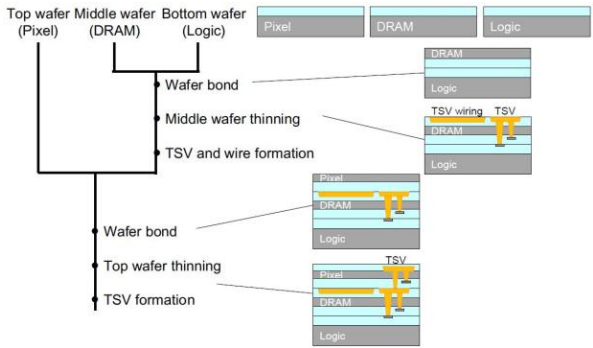

Fig. 10. Process flow of 3-layer stacked CIS using TSV [11]

### C. Experimental Verification using Test Module

Various tests were carried out to analyze the electrical and reliability characteristics of the TSV interconnects in the 3-layer stacked CMOS process. The authors fabricated a 300mm wafer with 9000 TSV chains that served as the test module. The TSV chain runs from the top of the pixel layer to the DRAM and then to the logic substrate. Fig. 11 shows the cumulative probability of the initial resistance measured for the 9000 TSV chains. Measuring resistance is critical for any type of interconnect as they are parasitic and create propagation delay for the signal flowing through them. From the results, the authors came to a conclusion that the initial resistance is small and there is little variation in the resistance value among different TSV chains. Further, the stress migration test was carried out on 1260 TSV units using the annealing process. During annealing, the test module was exposed to 175°C for 1000 hours and the plot in Fig. 12 depicts that the resistance shift is less than 2% for the 1260 units. The shift in resistance is calculated from before and after the annealing. In addition to that, the leakage current between the TSV and the DRAM substrate was measured by applying 1V between them. Leakage current is a critical factor when employing DRAM, as high leakage would lead to low retention memory time in DRAM. From the Fig. 13, the authors found that the leakage current is very low exhibiting small variation among the different units.

Dielectrics get damaged and form conducting paths when exposed to high electric field for a long time. Hence, the Time Dependent Dielectric Breakdown (TDDB) test was conducted on the test module by applying 180V between the DRAM and TSV for a prolonged time. After 10,000 seconds, the authors concluded that the electrical breakdown was considerably less implying that the insulating property is sufficient between the DRAM substrate and the TSVs and this is seen in Fig. 14.

This concludes the experimental tests done on the 300 mm test module.

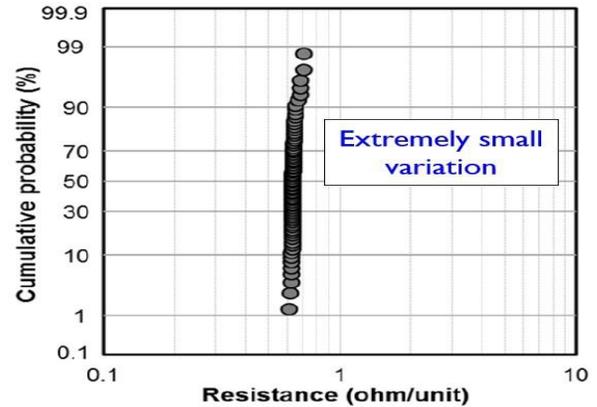

Fig. 11. Cumulative plot of TSV resistance [Modified [11]]

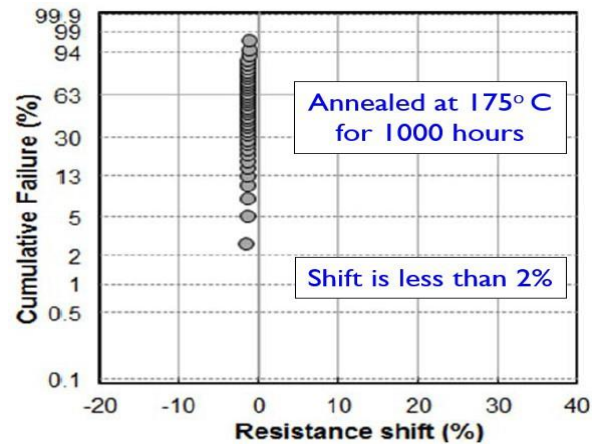

Fig. 12. TSV stress migration test [Modified [11]]

### D. Fabrication of the 3-layer pixel/DRAM/logic chip

Following the test module validation, a 3-layer pixel/DRAM/logic stacked chip was fabricated and its cross section is shown in the Fig.15. In the figure, the pixel shows the entire cross section and the peripheral shows the stacked TSV chain making the pixel-DRAM and DRAM-logic substrate connections. The upper TSVs are 2.5μm in diameter, connecting the pixel and DRAM substrates and there were about 15,000 of them in the stacked chip. The lower TSVs are about 3.5μm in diameter and there were 20,000 of them connecting the DRAM and logic substrates. The TSV pitch measured from the center of one TSV to another was 6.5μm.

Despite adding a third DRAM layer to the pixel/logic stack, the authors achieved the same thickness as the 2-layer chip without the DRAM. The substrates are usually thinned as part of the bonding process and the final thickness of DRAM substrate after thinning was 3μm. The fabricated pixel/DRAM/logic stacked BI-CMOS chip was 130μm in thickness with 19.3 megapixels and pixel size of 1.22μm x 1.22μm, yielding a 30fps at the output interface.

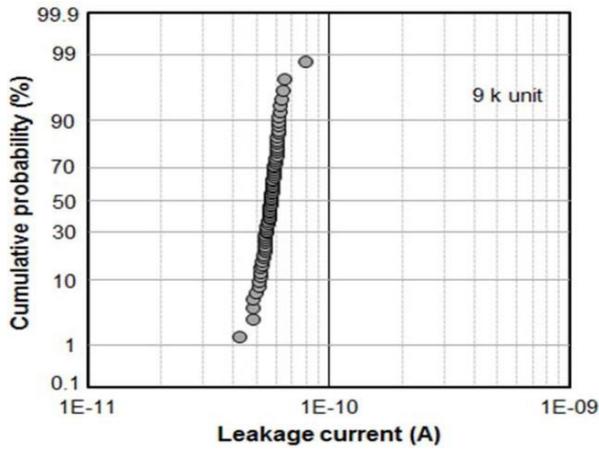

Fig. 13. TSV leakage current [11]

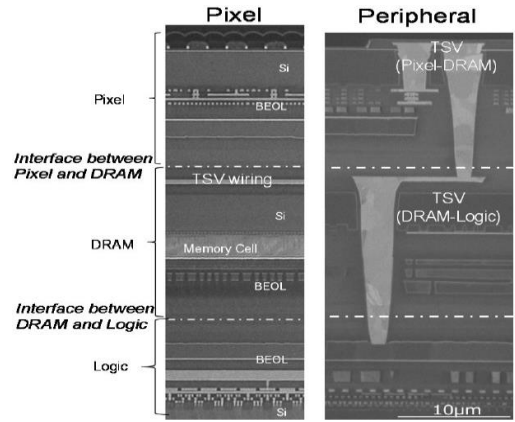

Fig. 15. Cross section of the fabricated 3-layer Stacked BI-CIS chip [11]

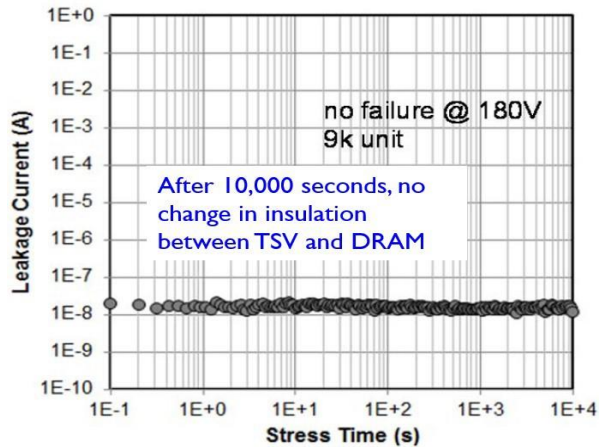

Fig. 14. TSV TDDB test [Modified [11]]

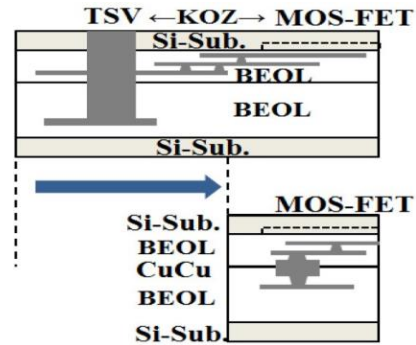

Fig. 16. Comparison of TSV (top) and Cu-Cu hybrid bonding (bottom) (CuCu showing significant reduction in chip size) [13]

### III. Stacked BI CIS using Cu-Cu hybrid bonding, Y. Kagawa et al, 2018

In [13], the authors have fabricated Stacked BI-CIS adopting the Cu-Cu hybrid bonding technology. TSV offers electrical connection between the substrates, whereas with hybrid bonding, the substrates are bonded electrically with Cu-Cu metal connection and physically by inter-layer dielectric (ILD) simultaneously. TSVs have few disadvantages compared to the Cu-Cu hybrid bonding as follows:

- TSVs require a special equipment for fabrication called deep silicon etcher.
- Fabrication of TSVs require Keep-Out-Zone (KOZ) around them, which is the clearance area that is needed to be void of any circuitry. This restricts the area available to the circuit designers for circuit design.
- TSV provides only electrical connection between the substrates.

A comparison of the TSV and Cu-Cu bonding is illustrated in the Fig. 16. It is noticeable that the Cu-Cu hybrid bonding technique requires less space and simplified fabrication process, making it the desired choice compared to TSVs. They also offer greater design flexibility for the circuit designers.

#### A. Process Flow of Cu-Cu hybrid bonding

The Cu-Cu bonding process [13], [14] begins with the parallel preparation of wafers Fig. 17(a). A thick dielectric layer is formed on the silicon using the chemical vapor deposition (CVD). CVD is the process of depositing a solid material in vapor form to achieve uniform thickness throughout the surface. Then, the trench and via which are part of the BEOL are made. Using the physical vapor deposition (PVD) method, copper seeds are formed in the trench. Following PVD, the trenches are filled with copper using the electro-chemical deposition (ECD). The excess copper is removed and very low dielectric roughness is attained by chemical mechanical polishing (CMP). Recessing of copper to a certain level is expected during CMP. As seen in 17 (b), the plasma activated wafers are brought together face-to-face and the dielectrics are bonded instantaneously. After CMP, annealing is done at 150°C to 300°C, due to which the metal expands to fill the gap between them. The aforementioned steps confirm that Cu-Cu hybrid bonding provides physical and electrical connections due to the dielectric and metal bonding between the substrates.

CMP plays a vital role in obtaining dielectric roughness and metal recess. In the standard Cu-Cu bonding process, the copper pads are recessed. [13] has adopted a controlled CMP process, where the copper pads are intentionally made to stick out and their precision to bond is controlled.

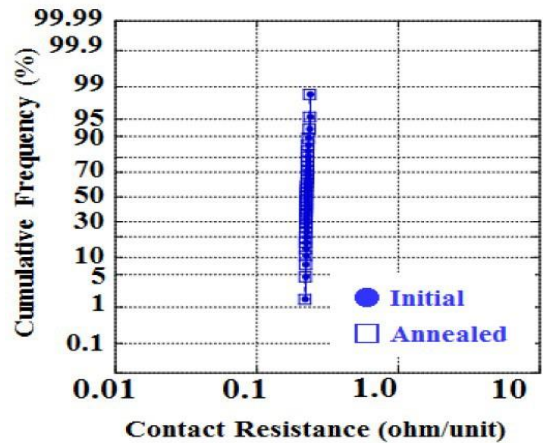

Fig. 18. Measured resistance of Cu-Cu interconnect [Modified [13]]

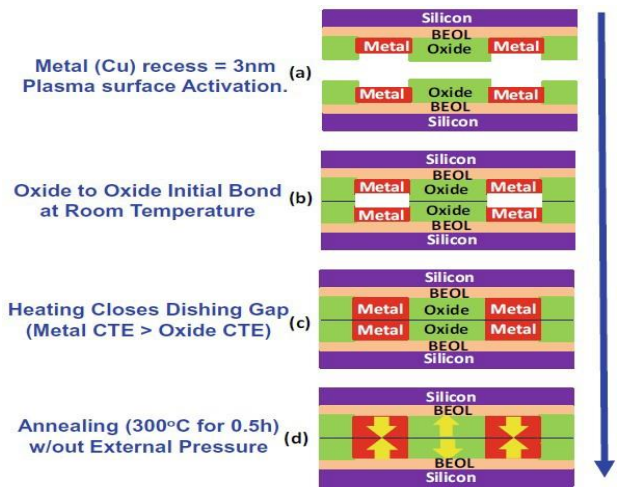

Fig. 17. Process Flow of Cu-Cu hybrid bonding [14]

*B. Experimental Verification using Test Module*

Similar to [11], the electrical and reliability tests for Cu-Cu bonding was carried on a 300mm test wafer. The 300mm wafer had 3 million Cu-Cu connections fabricated with $4\mu m$ pitch. In addition to measuring the resistance and contact pitch for the copper interconnects, the wafers were subjected to TDDB test to measure the chip lifetime. Fig. 18 and Fig. 19 respectively show the resistance and contact pitch measurements for the test module. Resistance measured before and after exposing the test wafer to 175°C for 1000 hours revealed that there was no significant change in the resistance value after the annealing process at a high temperature. A contact pitch of $4\mu m$ was achieved for the 3 million number of Cu-Cu connections in the test wafer. Fig. 19 illustrates that this work [13] has proven to have the minimum contact pitch compared to the previous works [15] and [16].

*C. Fabrication of the Stacked BI-CIS with Cu-Cu hybrid bonding*

The authors subsequently fabricated the stacked BI-CIS using Cu-Cu hybrid bonding technology. In the standard Cu-Cu bonding process, the copper pads recess during the CMP process, but the metal expands and the connection is made during annealing. But there is problem of bonding voids at the interface. Therefore, in [13], a controlled CMP process was introduced to ensure fine electrical connections between the copper pads of the substrates. In this specialized process, the copper pads were made to stick out intentionally due to which, the substrate-to-substrate metal connections made were robust without any bonding voids. Fig. 20 portrays the cross section of the stacked BI-CMOS from which it is apparent

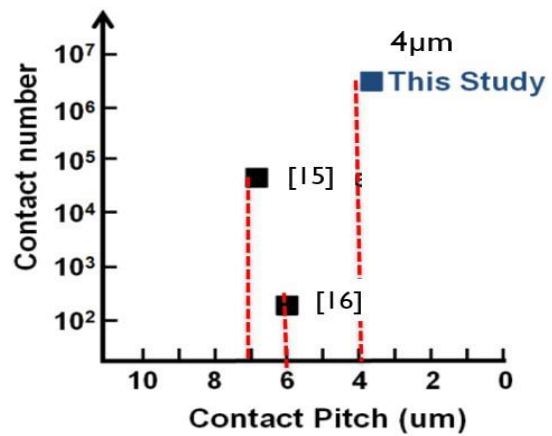

Fig. 19. Benchmark of number of contacts vs pitch [Modified [13]]

that the substrates are bonded with no voids at the bonding interface.

In the stacked BI-CMOS chip based on Cu-Cu hybrid bonding [13], a low contact pitch of $4\mu m$ was achieved, which played an important role in making small size image sensor chips with enhanced design flexibility. The fabricated chip had 22.5 megapixel resolution with $1\mu m$ x $1\mu m$ pixel size.

IV. Stacked 2-Layer BI CIS, K. Zaitsu et al, 2022

Industries have been working for decades on bringing out the best and optimized performances of the image sensors, while simultaneously working on reducing the chip size. In the previous CIS such as the [11] and [13], the photodiodes (PD) and the pixel transistors shared the same layer. Hence, the design area available to enhance either of their performance is restricted. Several efforts have been made to increase the CMOS image sensor's density for better performance. One such attempt [17], [18], [19] was to miniaturize pixels, but make deep PD and increase their fill factor. Despite that, photodiode capacity did not improve due to the ion implantation process which absorbs most of the photons. Another methodology [20] to improve the PD volume required long

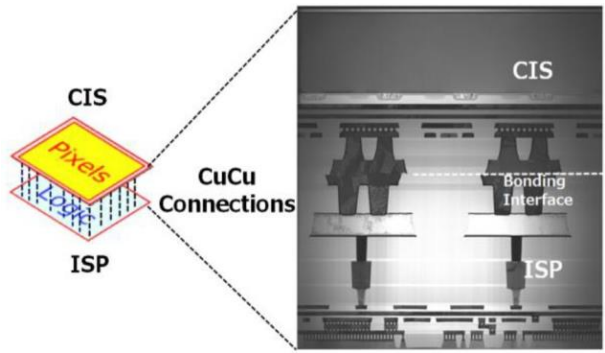

Fig. 20. Cross section of the Stacked BI-CIS using Cu-Cu hybrid bonding [13]

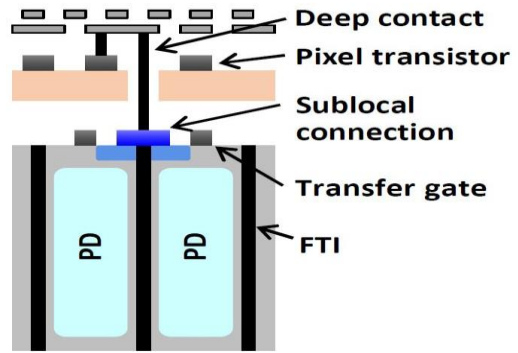

Fig. 22. Cross section of the 2-layer Stacked CIS [21]

wiring to connect the sensing nodes which led to decreased conversion gain.

### A. Structure of the 2-layer photodiode/pixel transistor

The research work in [21] introduces a 2-layer Stacked Back Illuminated CMOS image Sensor built using a 3-Dimensional sequential fabrication process. The 2-layer configuration of the photodiode/pixel transistor is illustrated in Fig. 21, where the photodiodes and the pixel transistors are designed independently in two different layers, thus creating opportunity for their optimized performance.

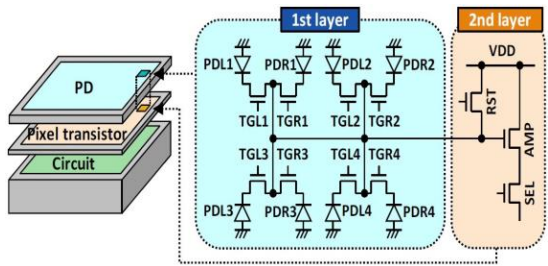

Fig. 21. 2-layer Stacked CIS architecture [21]

The 2-layer stacked CIS architecture can be better explained using the cross section shown in Fig. 22. The top layer has the photodiodes and transfer gates and the second layer has the amplifier, select gate transistors, reset gate transistors and deep contacts that connect the first and second layers. The photodiodes are isolated from each other by a process called Full Trench Isolation (FTI).

### B. Design Optimization

Few optimizations in the sequential fabrication process improved the CMOS image sensor's performance parameters such as Conversion Gain (CG), Quantum Efficiency(QE) and Full Well Capacity(FWC).

*1) Conversion Gain:* The pixel conversion gain (CG) is the voltage difference that the AMP (amplifier transistor) outputs for an electron charge from the photodiode to the Floating Diffusion(FD) node [22]. Fig. 23 illustrates the pictorial representation of conversion gain. Floating Diffusion is the sensing node capacitance at the gates of the RST and AMP (amplifier transistor). As part of the FTI process, numerous deep contacts are required to connect each of the photodiodes to its pixel transistor which increases the FD capacitance. Increase in capacitance decreases the conversion gain. Therefore, in an effort to minimize FD, groups of deep contacts are connected together to form a junction called sublocal connections as shown in Fig. 22. Sublocal connections diminish the floating diffusion, which in turn enhances the conversion gain.

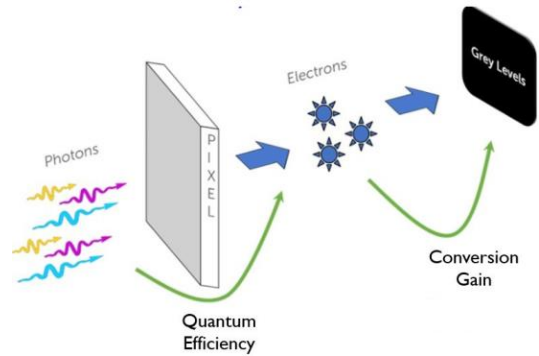

Fig. 23. Conversion gain and Quantum Efficiency [23]

*2) Full Well Capacity:* According to [24], "Full Well Capacity (FWC) is defined as the amount of charge that can be stored in the pixel before getting saturated". Fig.24 shows the full well capacity for different pixel size, which implies that higher FWC is achieved with more photodiode area. It is a common procedure in CIS to employ dual photodiodes (PD) - right (PDR) and left(PDL). Conventionally, boron impurities were used to isolate the dual PDs, which took up the photodiode volume, thereby decreasing the FWC. In the 2-layer architecture, partial FTI with reduced boron implantation is used to isolate the PDs.

*3) Quantum Efficiency:* As stated in [25], "Quantum efficiency (QE) is the measure of how effective an imaging device is at converting incident photons into electrons". This is depicted in Fig. 23 and are usually expressed in percentage. In other words, QE denotes the number of photons that are converted to electrons by the photodiode. In the conventional

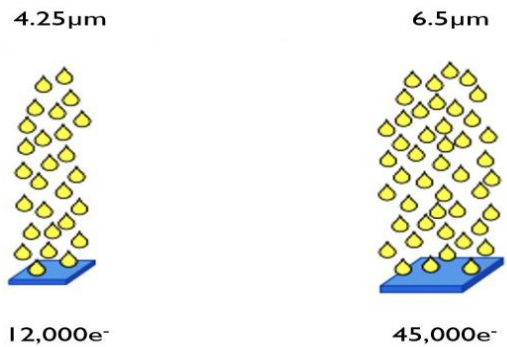

Fig. 24. Full Well Capacity [24]

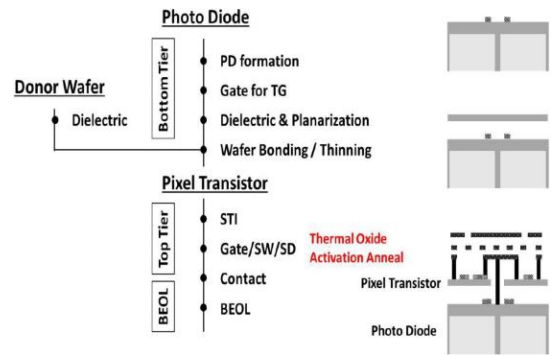

Fig. 25. Process Flow of 3D Sequential Integration [26]

FTI process, the trench is usually filled with poly-Silicon, but it absorbs portion of the light and lessens QE. For that reason, the authors utilized Silicon Oxide as the embedding material. Silicon oxide prohibits light from being absorbed in the trench and as a result, quantum efficiency is increased.

In summary, utilizing partial FTI process filled with Silicon oxide in the 2-layer architecture is beneficial for both QE and FWC.

### C. Process Flow of 3D Sequential Integration

3D sequential integration process works by building wafers layer by layer one on top of the other. The challenge in sequential integration is the fabrication of top layers without breaking down the substrates that are at the bottom due to the high temperature applied during the process [26]. The conventional parallelly processed chip fabrication [11], [13] involves front-end-of-line (FEOL) and back-end-of-line (BEOL) fabrication steps. FEOL is the first part of the fabrication process where the individual components such as transistors, resistors and capacitors are built and the interconnects between various layers are done in the final step, BEOL. In the sequential process, the deep contacts from the photodiodes and pixel transistor layers are connected by metal interconnects, which takes place between the top and bottom tiers.

The 3D sequantial integration is illustrated step-by-step in the Fig.25. Unlike the parallelly processed wafers, in the sequential process, the devices need not be prepatterned for bonding together. Instead, the top device tier is processed over the bottom tier device. As seen in the Fig. 25, the first step is to form the top tier devices, which is done by forming an active layer over the patterned wafer. This is the top tier FEOL, after which the interconnects between the two layers are etched and filled with metal to establish contact between the layers.

In [21], photodiodes and transfer gates are the bottom tier devices and the pixel transistors are the top tiers. To connect the photodiodes and the pixel transistors, the deep contacts from both these layers are connected through metal interconnects. During the annealing process, there is great possibility of the photodiodes and transfer gates being affected during the top tier formation. Hence, the authors chose an appropriate dielectric film that would provide thermal stability during the top tier fabrication.

### D. Fabrication of the 2-layer CIS and Optimized Results

The authors fabricated the 2-layer photodiode/pixel CIS using sequential integration process and measured the image sensor's performance parameters. A cross section of the 2 layer CIS is shown in the Fig. 26, which shows the location of the transfer gates, pixel transistors, photodiodes and deep contacts. The fabricated device has photodiodes of 1$\mu$m x 1$\mu$m in size. With the increased PD volume due to the partial FTI, the authors achieved high FWC of 12,000 e$^-$ compared to the previous works [27] [28] and this is evident in the Fig.27.

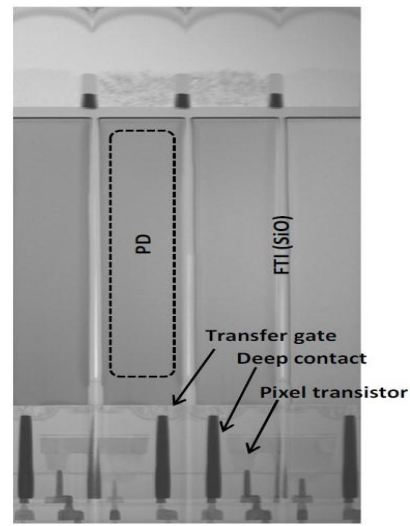

Fig. 26. Cross section of the 2-layer architecture [21]

As mentioned earlier, QE has improved with the combination of partial FTI and using silicon oxide as the filling material for the trench. QE increased by 19% at 530nm wavelength as seen in Fig. 28. Poly-silicon absorbs more light compared to the silicon oxide and this is demonstrated using the cross sectional simulation of the FTI in Fig. 29 and this is detrimental to QE.

As shown in the Fig.30, the conversion gain increased by 28% and the random noise decreased by 14% compared to the case without the sublocal connections of the deep contacts. The sublocal connections decrease the number of deep contacts and as a result reduce floating diffusion capacitance.

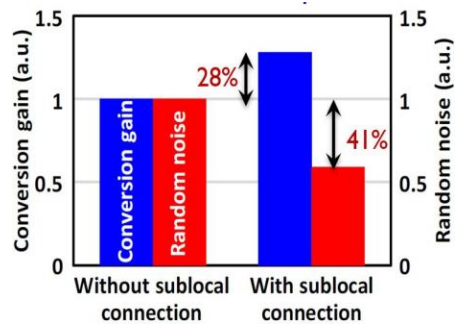

Fig. 30. Conversion Gain and Random Noise [Modified [21]]

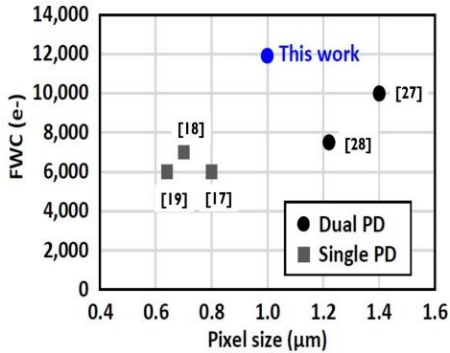

Fig. 27. Optimized FWC compared to previous works [Modified [21]]

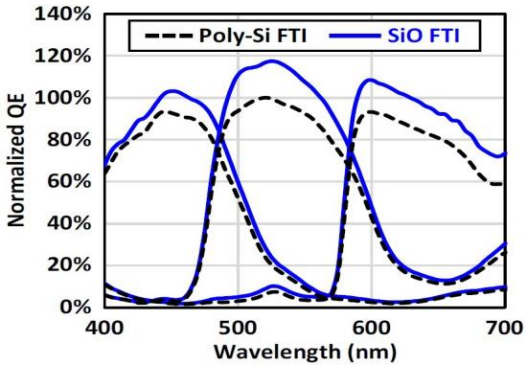

Fig. 28. Optimized QE - PolySilicon FTI vs Silicon Oxide FTI [21]

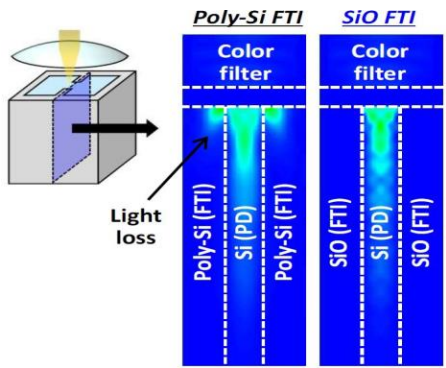

Fig. 29. Simulated cross section of light absorption for Poly-Si FTI (left) and SiO FTI (right) [21]

## V. Conclusion

According to [29], market share for CMOS Image sensors was USD 19.28 Billion in 2022 and it is expected to reach USD 38.78 Billion by 2030. This is mainly due to the increasing demand from smartphones, digital cameras and various new applications and being able to meet those demands. The design requirements from the various devices range from better image resolution, reduced image distortion to increased gain while obtaining small chip size at the same time. The research papers were reviewed here chronologically and it is evident that the fabrication processes played an important role in the transformation of the image sensors architecture and their performance.

Back Illumination and stacking layers are the major concepts that marked the beginning of the use of CMOS Image sensors in smartphones. Stacked layers helped with the reduction of chip size, while back illumination improved the sensor performance. TSV based fabrication process with the addition of DRAM layer played an important role in reducing the rolling shutter distortion and improving the image quality. Reading speed at the pixel side improved to 120 fps while maintaining the same reading speed, 30 fps at the output, ensuring that the delay is invisible to the consumer. Later, Cu-Cu hybrid bonding technology was adopted which proved to be an efficient and a convenient method for substrate bonding while successfully providing design flexibility and reduced chip size. This makes it possible to integrate more than one camera in the smartphones.

Finally, the 2-layer architecture was studied, where the pixel and photodiode are placed in two different layers using the 3D sequential integration process. In addition to adopting the process, other fabrication techniques such as sublocal connections and partial FTI with silicon oxide were employed through which high FWC of 12,000 e$^-$ was obtained, conversion gain increased to 28% and quantum efficiency increased by 19% compared to previous works. Additionally, the 2-layer architecture increased the dynamic range of the photodiodes as the different shades of the bright and dark areas are captured in the image, thereby improving the image quality.

It is anticipated that the CMOS image sensors will continue to be on the smartphones and have the potential to replace digital cameras in the distant future.